\documentclass[twocolumn,amsmath,amssymb,prb,longbibliography]{revtex4-1}
\usepackage{multirow}
\usepackage{array}
\usepackage{amsmath}
\usepackage{graphicx}
\usepackage{dcolumn}
\usepackage{bm}
\DeclareMathAlphabet \mathbfcal{OMS}{cmsy}{b}{n}

\begin{document}

\title{Femtosecond currents in transition metal dichalcogenides monolayers}

\author{S. Azar Oliaei Motlagh}
\author{Vadym Apalkov}
\author{Mark I. Stockman}
\affiliation{Center for Nano-Optics (CeNO) and
Department of Physics and Astronomy, Georgia State
University, Atlanta, Georgia 30303, USA
}

\date{\today}
\begin{abstract}
We theoretically study the interaction of an ultrafast intense linearly polarized optical pulse
with monolayers of transition metal dichalcogenides (TMDCs). Such a strong pulse redistributes
electrons between the bands and generates femtosecond currents during the pulse. Due to the large
bandwidth of the incident pulse, this process is completely off-resonant. While in TMDCs the
time-reversal symmetry is conserved, the inversion symmetry is broken and these monolayers have
the axial symmetry along armchair direction but not along the zigzag one. Therefore, the pulse
polarized along the asymmetric direction of TMDC monolayer generates both longitudinal, i.e., along
the direction of polarization, and transverse, i.e., in the perpendicular direction, currents. Such
currents result in charge transfer through the system. We study different TMDC materials and
show how the femtosecond transport in TMDC monolayers depend on their parameters, such as
lattice constant and bandgap.



 
\end{abstract}
\maketitle
\section{Introduction}

Nowadays, the femtosecond and strong field driven phenomena, e.g., high harmonic generations, the ultrafast ionization and metalization, the nonlinear current generations, and the nonlinear optical absorption in solids attract growing  interest due to their possible applications in ultrafast optical switches, optoelectronic devices and ultimately in ultrafast information processing\cite{Kiemle_etal_Nanophotonics_2020_ultrafast_vanderwaals,Yakovlev_2020_Nat_comm_dielectric,
Schiffrin_at_al_Nature_2012_Current_in_Dielectric,Ghimire_etal_Nat_2020_Comm_HHG, Stockman_et_al_Nat_Phot_2013_CEP_Detector,Apalkov_Stockman_PRB_2012_Strong_Field_Reflection,Ghimire_IOP_2020_HHG_crystal, Higuchi_Hommelhoff_et_al_Nature_2017_Currents_in_Graphene, Gruber_et_al_ncomms13948_2016_Ultrafast_pulses_graphene, Stockman_et_al_PhysRevB.95_2017_Crystalline_TI,Stockman_et_al_PhysRevB.98_2018_3D_TI, 
Hommelhoff_et_al_PhysRevLett.121_2018_Coherent,  Hommelhoff_et_al_1903.07558_2019_laser_pulses_graphene, sun_et_al_nnano.2011.243_2012_Ultrafast_pulses_graphene, Mashiko_et_al_Nature_Communications_2018_ultrafast_pulse_solid, Shin_et_al_IOP_Publishing_2018_ultrafast_pulse_solid,Hommelhoff_et_al_PhysRevLett.121_2018_Coherent, 
Gruber_et_al_ncomms13948_2016_Ultrafast_pulses_graphene, Higuchi_et_al_Nature_2017, Leitenstorfer_et_al_PhysRevB.92_2015_Ultrafast_Pseudospin_Dynamics_in_Graphene, Stockman_et_al_PhysRevB.98_2018_Rapid_Communication_Topological_Resonances, Sun_et_al_Chinese_Physics_B_2017_Ultrafast_pulses_TMDC, Zhang_et_al_OSA_2018_ultrafast_pulse_TMDC,Ghimire_et_al_Nature_Communications_2017_HHG, Reis_et_al_Nat_Phys_2017_HHG_from_2D_Crystals,
Simon_et_al_PRB_2000_Strong_Field_Fs_Ionization_of_Dielectrics,
Rosa_et_al_Optical_Materials_Express_2017_stacking_graphene_saturable_absorbers,
Kumar_et_al_APL_2009_saturable_absorption_graphene,
Gesuele_Photonics_2019_Transient_Absorption, Stockman_etal_PhysRevB.101.165433_Absorbance_gapped_graphene}. Among solids, transition metal dichalcogenides (TMDC) have unique optical and electrical properties.
The bulk TMDCs are stacks of monolayers, which are bounded by the van der Waals forces \cite{Strano_et_al_nnano.2012.193_Transitional_Metal, Novoselov_et_al_Science_2016_2D_Materials_and_Heterostructures}. Due to natural weakness of these forces, the bulk can be easily exfoliated to atomically thin monolayers \cite{Novoselov_et_al_Science_2016_2D_Materials_and_Heterostructures,Xiao_et_al_PRL_2012_Coupled_Spin_and_Valley_Physics}. Each monolayer is made of one layer of transition metal atoms like Mo and W, which is sandwiched between two chalcogin (S, Se, Te) layers. The monolyers can be found in different phases, while the semiconducting phase is the most common one. It has trigonal prismatic crystalline structures with $D_{3h}$ point symmetry group \cite{Liu_et_al_PRB_2014_Three_Band_Model}.  

The TMDC monolayers are direct bandgap semiconductors with the bandgaps of 1.1-2.1 eV\cite{Liu_et_al_PRB_2014_Three_Band_Model}. Similar to graphene, TMDC monolayers have honeycomb crystal structure but they are not centrosymmetric and the  inversion symmetry is broken. Due to the broken inversion symmetry, the Berry curvature is not singular but has finite values with opposite signs in two valleys, $K$ and $K^\prime$. The finite Berry curvature gives rise to an anomalous Hall effect in the absence of external magnetic field 
\cite{Nagaosa_Anomalous-Hall-effect_RevModPhys_2010}
Another difference of these materials from graphene is the excistence of strong intrinsic spin orbit coupling \cite{Liu_et_al_PRB_2014_Three_Band_Model}, which results in the spin splitting of the valence band (VB) and the conduction band (CB)\cite{Liu_et_al_PRB_2014_Three_Band_Model} and makes TMDC monolayers suitable for spintronic applications.

Previously, we have shown that a single cycle of a circularly polarized optical pulse induces a large valley polarization, $\eta_v \geq 40\%-60\%$, in TMDC monolayers, $\mathrm{MoS_2}$ and $\mathrm{WS_2}$  \cite{Stockman_et_al_PhysRevB.98_2018_Rapid_Communication_Topological_Resonances}.
The mechanism of producing fundamentally fastest valley polarization in these monolayers is independent of electron spin and has topological origin. Predominant population of one of the valleys in TMDC monolayer is not due to 
the optical selection rule as in case of a continuous wave but due to topological resonance, which is a competition of the dynamic phase and the topological phase that is accumulated during ultrashort and strong pulse \cite{Stockman_et_al_PhysRevB.98_2018_Rapid_Communication_Topological_Resonances}. 
It has been also recently predicted that the valley polarization can be tuned  by the bandgap in gapped graphene monolayers\cite{Stockman_et_al_PhysRevB.100.115431_2019_Gapped_Graphene}. In graphene, the inversion symmetry can be broken by placing graphene on different substrates, e.g., SiC, which reduces the point group symmetry of graphene from  $D_{6h}$ to $D_{3h}$ \cite{Lanzara_et_al_Nat_Mat_2007_Gapped_Graphene,Conrad_et_al_PhysRevLett.115_2015_Gapped_Graphene_on_SiC}. 

In the field of intense optical pulse the valence and conduction band states are strongly coupled, which results in 
generation of strong nonlinear electric currents and finite transfer of electric charge through the system. Thus ultrafast optical pulses allow to control the transport properties of electron systems and enhance the conductivity of solids on the femtosecond time scale. Understanding of the extent of such control is important for possible device application of different solids.  
In this article we study the femtosecond currents driven by a single-cycle of an intensive laser pulse in monolayers of different TMDC materials. Different characteristics, e.g., the energy dispersion and the lattice constants, of these materials strongly affect the generated electric current and correspondingly transferred charge. The generated electric current also depends on the direction of polarization of the optical pulse\cite{Stockman_et_al_J.Phys.Condens.Matter_2019_current_gapped_graphene}.



\section{Main Equations}

We consider coherent ultrafast electron dynamics in the field of the pulse, assuming that the 
relaxation and scattering times in TMDC monolayers are longer than 10 fs \cite{Hwang_Das_Sarma_PRB_2008_Graphene_Relaxation_Time,
Breusing_et_al_Ultrafast-nonequilibrium-carrier-dynamics_PRB_2011,Ultrafast_collinear_scattering_graphene_nat_comm_2013,Gierz_Snapshots-non-equilibrium-Dirac_Nat-Material_2013,Nonequilibrium_dynamics_photoexcited_electrons_graphene_PRB_2013}.
The time dependent Hamiltonian of the system has the following form 
\begin{equation}
H(t)=H_0-e\mathbf r\mathbf F(t), 
\label{H}
\end{equation} 
where $H_0$ is the field-free Hamiltonian of a TMDC monolayer, $e$ is an electron charge, $\bf{r}$ is the position vector, and  $\mathbf F(t)$ is the electric field of the pulse. We consider the three-band tight binding model for TMDC monolayer, which gives three bands: one valence band and two conduction bands.  
For the pulse, linearly polarized in the $x$-direction, the electric field is given by the following expression 
\begin{equation}
F_x(t)= F_0(1-2u^2)e^{-u^2}~,~~~F_y(t)=0,
\label{Field}
\end{equation}
where $u=t/\tau$, and $\tau=1$ fs is the pulse duration.

\begin{figure}
\begin{center}\includegraphics[width=0.47\textwidth]{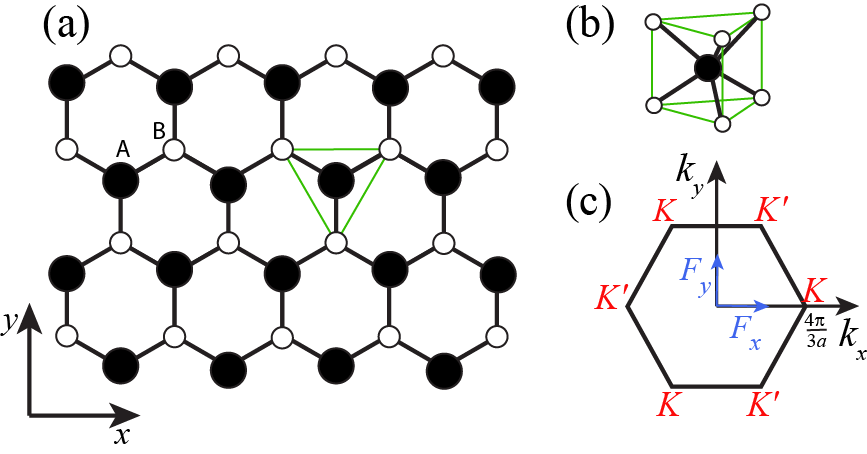}\end{center}
\caption{(Color online) Crystal structure of TMDC monolayer. (a,b) Honeycomb crystal structure of TMDC monolayer consists of two 
sublattices: $A$ and $B$. Sublattice A is occupied by transition metal atoms (closed dots), while sublatice B is occupied by chalcogen atoms (open dots). (c) The first Brillouin zone with two valleys, $K$ and $K^{\prime }$. For general polarization of the optical pulse, the electric field in the pulse has both $x$ and $y$ components, $F_x$ and $F_y$. 
} 
\label{TMDC_Lattice}
\end{figure}

The electron dynamics is determined by the solution of the corresponding time dependent Schrodinger equation (TDSE)
\begin{equation}
i\hbar \frac{d\Psi (t) }{dt } = H(t) \Psi(t) .
\label{TDSE}
\end{equation}
It is convenient to express this solution in the basis of time dependent Houston functions \cite{Houston_PR_1940_Electron_Acceleration_in_Lattice} 
\begin{equation}
\Phi^{(H)}_{\alpha {\bf q}}({\bf r},t)=\Psi^{(\alpha)}_{{\bf k}({\bf q},t)}({\bf{r}})\exp{(i\phi_\alpha^{(D)}(q,t)+i\phi_\alpha^{(B)}(q,t))}~,
\end{equation}
where $\Psi^{(\alpha)}_{{\bf k}}({\bf{r}})$ are the eigenfunctions of the time independent part of  Hamiltonian $H_0$ and $ \phi_\alpha^{(D)}(q,t)={- \frac{1}{\hbar}} {\int} {dt^\prime E_\alpha[{\bf k }({\bf q},t^\prime)]}$ is the dynamic phase, $E_\alpha $ are the eigenvalues of $H_0$ , $ {\phi_\alpha^{(B)}(q,t)={- \frac{e}{\hbar}} {\int} {dt^{\prime} \mathbf{F}(t^\prime)\mathbf{A}_{\alpha\alpha}[{\bf k }({\bf q},t^\prime)]}}$ is the Berry phase, $\mathbf{A}_{\alpha \alpha }$ is the Berry connection, which is defined below by Eq. (\ref{D}), and $ {\alpha \in \lbrace v,c_1,c_2 \rbrace} $ where $v,c_1,c_2$ denote the  VB and two  CBs, respectively. The electron trajectory in the reciprocal space, $\mathbf k(\mathbf q, t)$, is given by the Bloch acceleration theorem \cite{Bloch_Z_Phys_1929_Functions_Oscillations_in_Crystals}, 
\begin{equation}
\mathbf k(\mathbf q, t)=\mathbf q+\frac{e}{\hbar}\int_{-\infty}^t \mathbf F(t^\prime) dt^\prime
\end{equation}
where $\mathbf q$ is the initial crystal wave vector.

In the basis of Houston functions, solutions of the time dependent Schrodinger equation (\ref{TDSE}) are parameterized 
by initial crystal wave vector $\mathbf q$ and are given by the following expression
\begin{equation}
\Psi_{\bf q} ({\bf r},t)=\sum_{\alpha=c_1,c_2,v}\beta_{\alpha{\bf q}}(t) \Phi^\mathrm{(H)}_{\alpha {\bf q}}({\bf r},t),
\end{equation}
where $\beta_{\alpha{\bf q}}(t)$ are expansion coefficients, which satisfy  the following system of differential equations 
\begin{equation}
i\hbar\frac{\partial B_\mathbf q(t)}{\partial t}= H^\prime(\mathbf q,t){B_\mathbf q}(t)~.
\end{equation}
Here
\begin{eqnarray}
B_\mathbf q(t)&=&\begin{bmatrix}\beta_{c_2\mathbf q}(t)\\\beta_{c_1\mathbf q}(t) \\ \beta_{v\mathbf q}(t)\\ \end{bmatrix}~,\\ 
H^\prime(\mathbf q,t)&=&-e\mathbf F(t)\hat{\mathbfcal A}(\mathbf q,t)~,\\
\hat{\mathbfcal{A}}(\mathbf q,t)&=&\begin{bmatrix}0&\mathbfcal{D}_\mathrm{c_2c_1}(\mathbf q,t)&\mathbfcal{D}_\mathrm{c_2v}(\mathbf q,t)\\
\mathbfcal D_\mathrm{c_2c_1}^{\ast}(\mathbf q,t)&0&\mathbfcal D_\mathrm{c_1v}(\mathbf q,t)\\
\mathbfcal D_\mathrm{c_2v}^{\ast}(\mathbf q,t)&\mathbfcal D_\mathrm{c_1v}^{\ast}(\mathbf q,t)&0\\
\end{bmatrix}.
\end{eqnarray}
where
\begin{eqnarray}
&&\mathbfcal D_{\alpha \alpha_1}(\mathbf q,t)=
\mathbfcal A_{\alpha \alpha_1}[\mathbf k (\mathbf q,t)]\times \nonumber \\ 
&&\exp\left(i\phi^\mathrm{(D)}_{\alpha\alpha_1}(\mathbf q,t)+i\phi^\mathrm{(B)}_{\alpha\alpha_1}(\mathbf q,)t\right),
 \label{D}
\\
&&\phi^\mathrm{(D)}_{\alpha\alpha_1}(\mathbf q,t)=\phi^\mathrm{(D)}_{\alpha_1}(\mathbf q,t)-\phi^\mathrm{(D)}_{\alpha}(\mathbf q,t),
 \label{phi_D}
 \\ 
 &&\phi^\mathrm{(B)}_{\alpha\alpha_1}(\mathbf q,t)=\phi^\mathrm{(B)}_{\alpha_1}(\mathbf q,t)-\phi^\mathrm{(B)}_{\alpha}(\mathbf q,t),
 \label{phi_B}
 \\ 
&&\mathbf D_{\alpha\alpha_1}=e \mathbfcal{A}_{\alpha\alpha_1}; ~
{\mathbfcal{A}}_{\alpha \alpha_1}({\mathbf q})=
\left\langle \Psi^{(\alpha)}_\mathbf q  |   i\frac{\partial}{\partial\mathbf q}|\Psi^{(\alpha_1)}_\mathbf q   \right\rangle .
\label{D}
\end{eqnarray} 
Here, ${\mathbfcal A}_{\alpha \alpha_1}(\mathbf k)$ is non-Abelian Berry connection \cite{Wiczek_Zee_PhysRevLett.52_1984_Nonabelian_Berry_Phase, Xiao_Niu_RevModPhys.82_2010_Berry_Phase_in_Electronic_Properties, Yang_Liu_PhysRevB.90_2014_Non-Abelian_Berry_Curvature_and_Nonlinear_Optics}, and $\mathbf D_{\alpha\alpha_1}$ is the interband dipole matrix, which determines the optical transitions between the VB and CBs. 

The crystal structure of TMDC monolayer is shown in Fig.\ \ref{TMDC_Lattice}. It has $D_{3h}$ symmetry and 
consists of two sublattices $A$ and $B$, which are occupied by transition metal atoms (sublattice A) and chalcogen atoms (sublattice B). The first Brillouin zone of TMDC monolayer is a hexagon with two 
valleys, $K$ and $K^{\prime }$ - see Fig.\ \ref{TMDC_Lattice}(c).   
We describe TMDC monolayer within the three band tight binding model\cite{Liu_et_al_PRB_2014_Three_Band_Model}. In this model only the couplings between the nearest neighbor $d$ orbitals ($d_{xy}$, $d_{z^2}$, and $d_{x^2-y^2}$) of transition metal atoms are considered. The corresponding Hamiltonian is the sum  of the nearest neighbor tight-binding Hamiltonian  $H^\mathrm{(TNN)}$, and  spin orbit coupling (SOC) contribution $H^\mathrm{(SOC)}$\cite{Liu_et_al_PRB_2014_Three_Band_Model},
\begin{eqnarray}
{H_0(\mathbf{k})}&=& I \otimes H^\mathrm{(TNN)}+H^\mathrm{(SOC)}\nonumber\\&=& \left[ {\begin{array}{cc}
{H^\mathrm{(TNN)}(\mathbf{k})+\frac{\mathrm{\lambda}}{2}{L_z}}&0\\0&{H^\mathrm{(TNN)}(\mathbf{k})-\frac{\mathrm{\lambda}}{2}{L_z}}
\end{array} } \right]\nonumber\\&=& \left[ {\begin{array}{cc}
{H^\mathrm{\uparrow}_{3\times3}(\mathbf{k})}&0\\0&{H^\mathrm{\downarrow}_{3\times3}(\mathbf{k})}
\end{array} } \right]~,
\label{eq:Total Hamiltonian}
\end{eqnarray}
where the tight binding matrix $H^\mathrm{(TNN)}$ is given in Appendix A,  $\lambda$ is the SOC constant\cite{Liu_et_al_PRB_2014_Three_Band_Model}, and
\begin{equation}
{L_z=}\left[ {\begin{array}{ccc}
{0} & 0 & 0\\
0 & 0 & {2i}\\
0 & {-2i} & 0
\end{array} } \right].
\end{equation}                                                                                                                                                                             
Since, the two spin components are not coupled by external electric field, we solve TDSE for each spin component independently.

The main parameters of TMDC monolayes, which are the bandgap, lattice constant, and SOC constant, are shown in Table I. The lattice constant is in the range of $3.19-3.56$ $\mathrm{\AA}$, while the bandgap lies between 0.8 eV and 2.0 eV.

\begin{table}
\begin{center}
\begin{tabular}{|c|c|c|c|c|c|c| }
\hline
 \\[-2mm]
\multirow{1}{6em}{}&$\mathrm{MoS_2}$ & $\mathrm{WS_2}$ & $\mathrm{MoSe_2}$ & $\mathrm{WSe_2}$ &$\mathrm{MoTe_2}$ & $\mathrm{WTe_2}$
 \\[-2mm]\\
\hline
\\[-2mm]
\multirow{1}{6em}{$\mathrm{a~(\AA)}$} &3.19&3.191&3.326&3.325&3.557&3.560 
 \\[-2mm]
 \\
\hline
\\[-2mm]
\multirow{1}{6em}{$\mathrm{\lambda~(eV)}$} &0.073&0.211& 0.091& 0.228&0.107& 0.237
\\[-2mm]
\\
 \hline
 \\[-2mm]
 \multirow{1}{8.5em}{$\Delta^{\mathrm{Up}}_K =\Delta^{\mathrm{Down}}_{K^{\prime}}$ (eV)} &1.590&1.600& 1.346& 1.325&0.967& 0.835
  \\[-2mm]
  \\
 \hline
 \\[-2mm]
  \multirow{1}{8.5em}{$\Delta^{\mathrm{Down}}_K =\Delta^{\mathrm{Up}}_{K^{\prime}}$ (eV)} &1.736&2.023& 1.526& 1.776&1.180& 1.307
   \\[-2mm]
   \\
 \hline
\end{tabular}

\end{center}
\caption{Lattice constant, spin orbit coupling constant, and bandgap (for spin up and spin down) at the $K$ and $K^\prime$ points  for different TMDC monolayers \cite{Liu_et_al_PRB_2014_Three_Band_Model}.}
\label{T1}
\end{table}

The femtosecond field-driven currents in solids generally have two main contributions, which come from interband and inraband dynamics. While these contributions are not gauge invariant quantities, the total current, which is the sum of two contributions, is gauge invariant\cite{Ernotte_PhysRevB.98.235202_2018_current_gauge_invariant}. We use the following expressions to calculate the intraband, $\mathbf J_\text{ra}$, and interband, $\mathbf J_\text{er}$, currents, 
\begin{equation}
\mathbf J_\text{ra}(t)  =\frac{e}{a^2}\sum\limits_{g_s}\sum\limits_{\alpha=\mathrm{v,c_1,c_2},\mathbf q}\left| \beta _{\alpha,g_s}(\mathbf q,t) \right|^2\mathbf v_{\mathrm{\alpha },g_s}{(\mathbf k(\mathbf q,t))}~,
\label{intra}
\end{equation}
where $\mathbf v_{\mathrm{\alpha},g_s}(\mathbf  k)=\frac{\partial}{\partial\mathbf k}E_{\mathrm{\alpha},g_s}(\mathbf k)$ is the group velocity (intraband velocity) and $g_s$ is the electron spin; the interband current is given by the following expression 
\begin{eqnarray}
  && \mathbf J_\text{er}(t)=i\frac{e}{\hbar a^2}\sum\limits_{g_s}\sum _{\substack{\mathbf q\\ \alpha,\alpha^\prime=\mathrm{v,c_1, c_2}\\
 \alpha\ne\alpha^\prime}}\beta _{\alpha^\prime,g_s}^\ast(\mathbf q,t)\beta _{\alpha,g_s}(\mathbf q,t)\nonumber \\&&\times\exp \{ i \phi^\mathrm{(D)}_\mathrm{\alpha^\prime\alpha,g_s}(\mathbf q,t)+ i \phi^\mathrm{(B)}_\mathrm{\alpha^\prime\alpha,g_s}(\mathbf q,t)\}\nonumber \\ 
 &&\times\left[ E_{\alpha^\prime,g_s}\left(\mathbf k(\mathbf q,t)\right)-E_{\alpha,g_s} \left(\mathbf k(\mathbf q,t)\right)\right] \mathbfcal A_{\alpha^\prime \alpha,g_s}\left(\mathbf k(\mathbf q,t)\right).\nonumber \\ 
 \label{J}
\end{eqnarray}


\section{Results and discussion}

Below we consider the following TMDC materials:  $\mathrm{MoS_2}$, $\mathrm{WS_2}$, $\mathrm{MoSe_2}$, $\mathrm{WSe_2}$, $\mathrm{MoTe_2}$, and $\mathrm{WTe_2}$. The parameters for these materials are taken from Ref.  \cite{Liu_et_al_PRB_2014_Three_Band_Model}. The crystal structure of the corresponding TMDC monolayer with the first  Brillouin zone is shown Fig. \ref{TMDC_Lattice}. Within the three band tight binding model the monolayer has one valence band (VB) and two conduction bands (CBs). Initially, i.e., before the pulse, the valence band is occupied and the conduction bands are empty. We apply a linearly polarized pulse propagating along $z$ direction with the amplitude of  $\sim 0.1 -0.5~ \mathrm{ V\AA^{-1}}$ and the duration of $\sim 5 ~\mathrm{fs}$. 

One the characteristics of electron dynamics in the field of the pulse is CB population distribution in the reciprocal space, $N_\mathrm{CB}\bf(k) = |\beta_{C_1,{\bf k}}|^2 + |\beta_{C_2,{\bf k}}|^2$. Such distribution is nonzero 
during the pulse and its residual value,  $N_\mathrm{CB}^\mathrm{(res)}\bf(k)$, determines irreversibility of the electron dynamics. As theoretical and experimental studies show the ultrafast electron dynamics  is irreversible in semimetals, e.g. graphene\cite{Hommelhoff_et_al_1903.07558_2019_laser_pulses_graphene,
Stockman_et_al_PRB_2016_Graphene_in_Ultrafast_Field}, Weyl semimetals \cite{Stockman_et_al_PhysRevB.99_2019_Weyl}and semiconductors, black phosphorene \cite{Stockman_et_al_PhysRevB.97.035407_2018_Phosphorene}, and TMDCs monolayers \cite{Hommelhoff_etal_2020_electron_dynamic_graphene_TMDC,Stockman_et_al_PhysRevB.98_2018_Rapid_Communication_Topological_Resonances}.
In addition to the irreversibility, the residual CB population distribution, $N_\mathrm{CB}^\mathrm{(res)}\bf(k)$, also determines the valley polarization after a circularly polarized pulse\cite{Stockman_et_al_PhysRevB.98_2018_Rapid_Communication_Topological_Resonances}.

.


\begin{figure}
\begin{center}\includegraphics[width=0.47\textwidth]{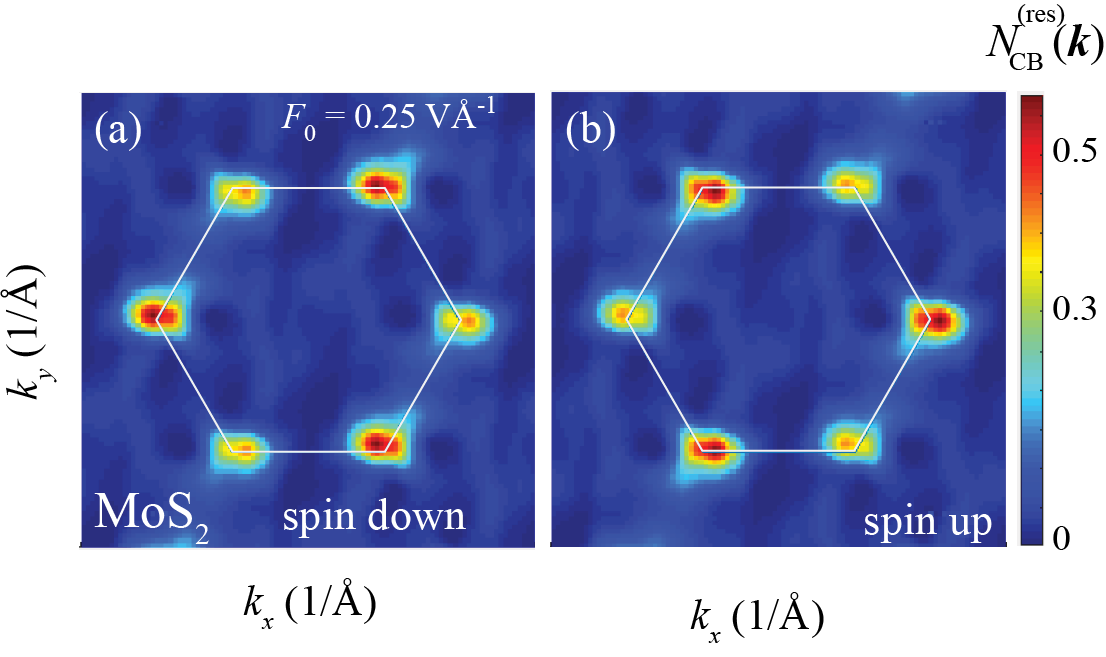}\end{center}
\caption{(Color online) Residual CB population distribution in the reciprocal space. The distribution is shown for $\mathrm{MoS_2}$ monolayer and for (a) spin down and (b) spin up components. The white solid lines show the boundary of the first Brillouin zone. The pulse is polarized in the $x$ direction. 
} 
\label{MoS2_Fx_0p25_Spin_down_up}
\end{figure}

\begin{figure}
\begin{center}\includegraphics[width=0.47\textwidth]{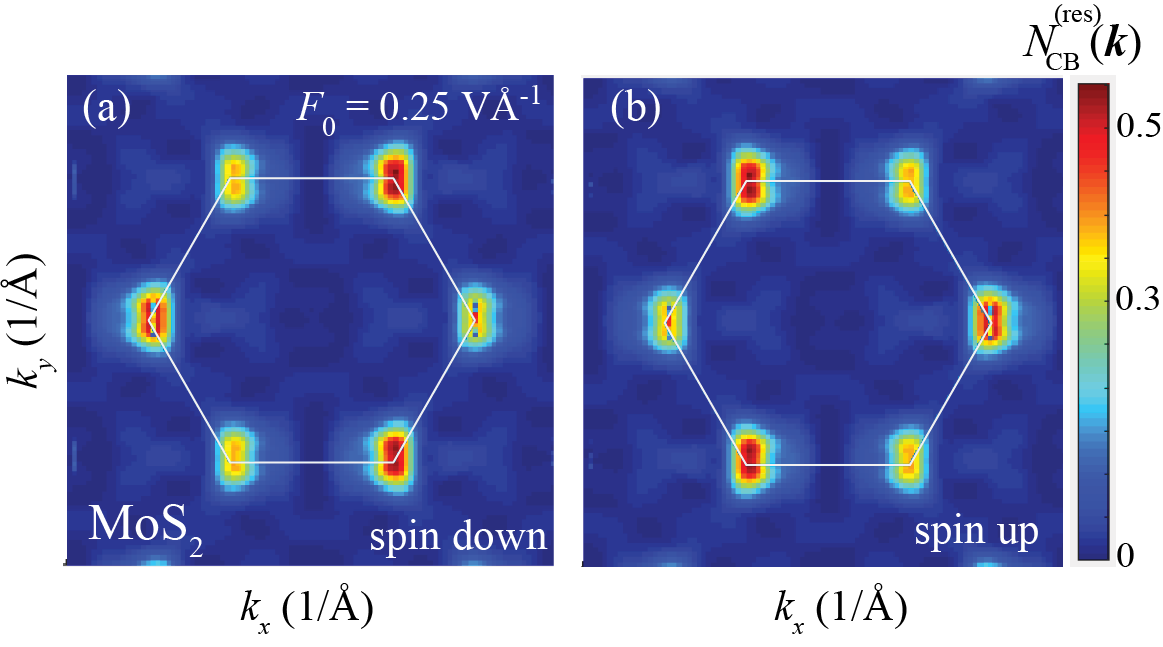}\end{center}
\caption{(Color online) The same as Fig. \ \ref{MoS2_Fx_0p25_Spin_down_up} but for the pulse polarized in the $y$ direction.
} 
\label{MoS2_Fy_0p25_Spin_down_up}
\end{figure}

Typical for TMDC monolayers, the residual CB population distribution in the reciprocal space  is shown in Fig.\ \ref{MoS2_Fx_0p25_Spin_down_up} for two spin components, down (a) and up (b). The pulse is linearly polarized in $x$-direction with the amplitude of $0.25~\mathrm{V\AA^{-1}}$.
The CB population is large near the $K$ and $K^\prime$ valleys, which is due to large interband coupling at these two points. For such small field amplitude, the population distribution does not show any interference fringes. 
For a given spin component, up or down, one valley is more populated than another one. For example, for spin down (see Fig.\ \ref{MoS2_Fx_0p25_Spin_down_up}(a)), the CB population of $K^\prime$ valley is higher than the corresponding population of the $K$ valley. However, the total CB population, summed over both spin components, is the same for 
both valleys. This is because the linear polarized pulse preserves the time reversal symmetry and does not induce any valley polarization. The axis $x$ is not the axis of symmetry of TMDC monolayer and the residual CB population distribution, shown in Fig.\ \ref{MoS2_Fx_0p25_Spin_down_up}, clearly shows such asymmetry. Because the CB population distribution is not symmetric with respect to the $x$ axis, the electric current is generated in 
both $x$ and $y$ directions.


The CB population distribution for the applied pulse polarized in $y$ direction is shown in Fig.\ \ref{MoS2_Fy_0p25_Spin_down_up}. Similar to the $x$ polarized pulse, the CB population is concentrated near the $K$ and $K^\prime$ valleys with zero residual valley polarization. The $y$ axis is the axis of symmetry of the system and 
the CB population distribution is symmetric with respect to the $y$ axis. Because of this symmetry, the electric current is generated during the pulse only in $y$ direction.

\begin{figure}
\begin{center}\includegraphics[width=0.47\textwidth]{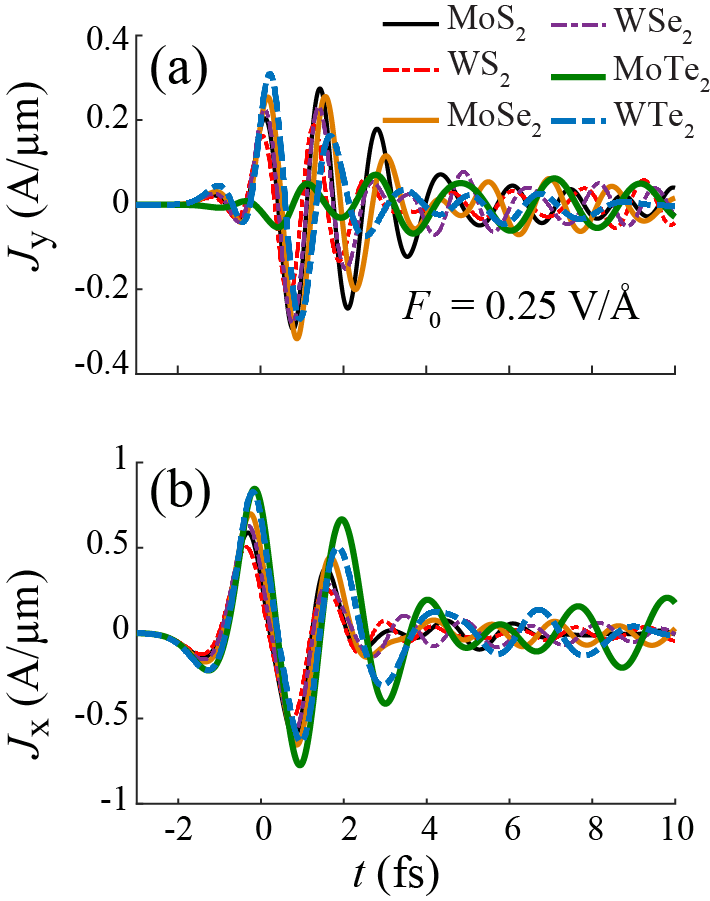}
\end{center}
\caption{(Color online) Femtosecond field driven currents as a function of time in different TMDC  monolayers. The generated electric currents have both the $y$ components (a) and the $x$ components (b). The pulse is linearly polarized in the $x$ direction and its  amplitude is $0.25~\mathrm{V\AA^{-1}}$.
} 
\label{TMDC_J_Fx_0p25_Linear}
\end{figure}

The ultrafast field driven intraband and interband electron dynamics generates an electric current. For the pulse polarized in $x$ direction, i.e., along the zigzag direction, the electric current is generated in both $x$ and $y$ directions\cite{Stockman_et_al_J.Phys.Condens.Matter_2019_current_gapped_graphene}. The current along $y$ direction, i.e., the direction perpendicular to the polarization of the pulse, strongly depends on the bandgap of TMDC monolayer. It disappears at zero bandgap, e.g., for pristine graphene, when the $x$ axis is the axis of symmetry. For the pulse polarized in $y$ axis, which is the axis of symmetry of TMDC monolayer, the electric current is generated only along the direction of polarization of the pulse. Below we consider only the electric pulse polarized in $x$ direction, which covers electron transport both in the direction of the pulse polarization and in the perpendicular direction. 


The generated electric currents for different TMDC materials are shown in Fig.\ \ref{TMDC_J_Fx_0p25_Linear}. The pulse is polarized in the $x$ direction so both $J_x$ and $J_y$ components of the current are nonzero. The field amplitude is $0.25$ V/\AA. 
The $x$ component of the current for all TMDC materials shows the same profile during the pulse, i.e, $-2 fs < t < 2 fs$, but after the pulse, $J_x$ has oscillatory behavior with the frequency of oscillations that depends on the bandgap of TMDC monolayer, which is in the range of $0.8$ - $2$ eV for the TMDC materials shown in the figure. Such oscillations in the residual current $J_x$ is due to the fact that the main contribution to the current is the interband one, while the intraband contribution, which depends only on the population of the conduction and valence bands and thus do not show oscillation after the pulse, is small.  

The generated current in the $y$ direction is almost three times smaller than the current in the $x$ direction. It also shows the oscillatory behavior as a function of time with well pronounced bandgap-dependent oscillations after the pulse. 
Although the profile of current $J_y$ during the pulse ($-2 fs<t< 2 fs$) is almost the same for all TMDC monolayers, one TMDC material, namely $\mathrm{MoTe_2}$, shows completely different time dependence.

\begin{figure*}
\begin{center}\includegraphics[width=1\textwidth]{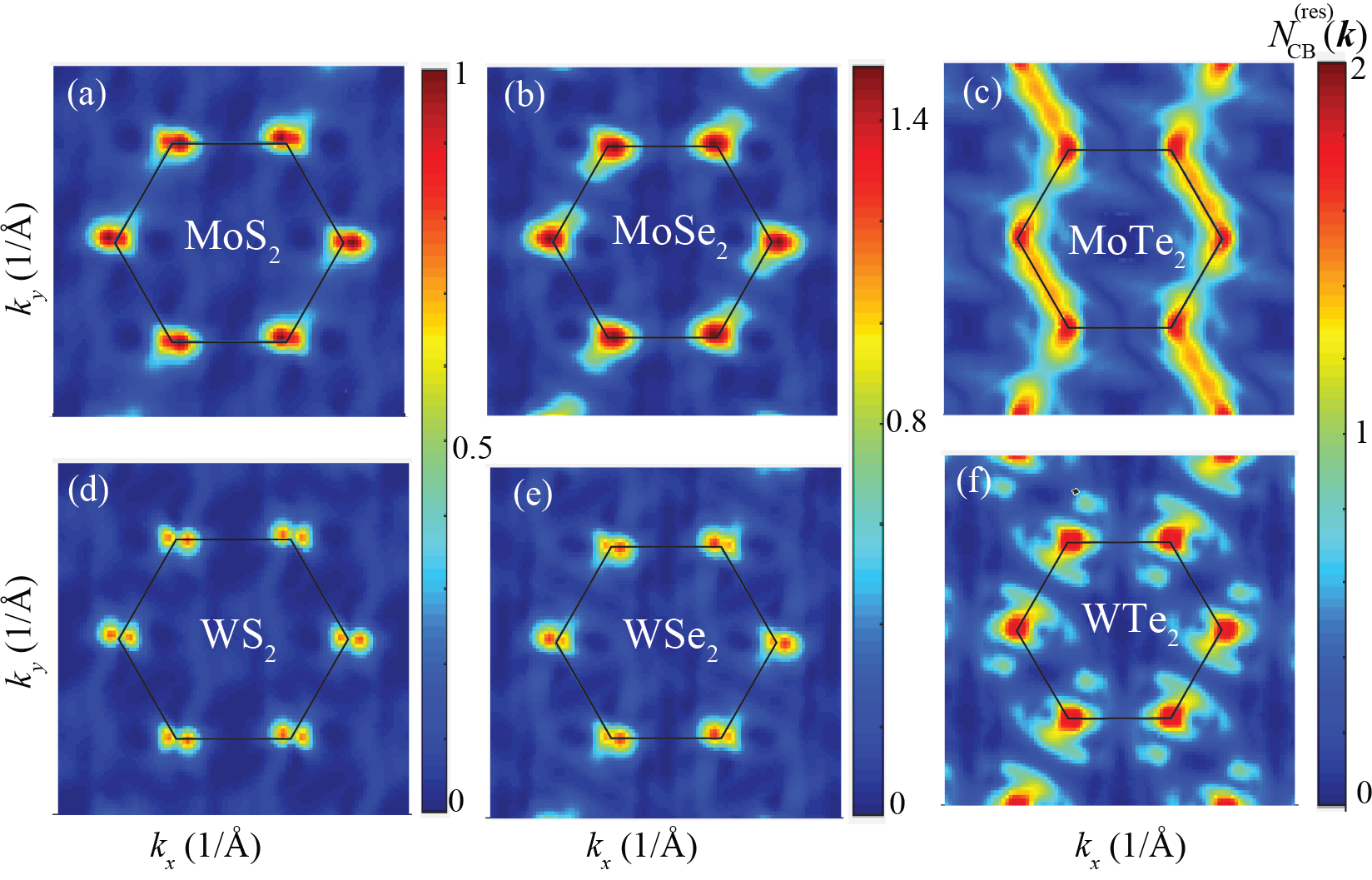}\end{center}
\caption{(Color online) Residual CB population distributions in the reciprocal space 
for different TMDC  monolayers: (a) $\mathrm{MoS_2}$, (b) $\mathrm{MoSe_2}$,  (c) $\mathrm{MoTe_2}$, (d) $\mathrm{WS_2}$, (e) $\mathrm{WSe_2}$, and (f) $\mathrm{WTe_2}$.  The optical pulse is linearly polarized in the $x$ direction and its  amplitude is $0.25~\mathrm{V\AA^{-1}}$. 
The black solid lines show the edges of the first Brillouin zone. For all TMDC monolayers, except $\mathrm{MoTe_2}$, the CB population is concentrated near the $K$ and $K^{\prime }$ points. 
} 
\label{Res_CB_TMDCs_Fx_0p25}
\end{figure*}

The unique behavior of $\mathrm{MoTe_2}$  monolayer can be understood from the corresponding CB population distribution in the reciprocal space. The residual CB populations are shown in Fig. \ref{Res_CB_TMDCs_Fx_0p25} for different TMDC monolayers. For all monolayers except $\mathrm{MoTe_2}$, $N_\mathrm{CB}^\mathrm{(res)}\bf(k)$ is concentrated at the $K$ and $K^\prime$ points along both $k_x$ and $k_y$ directions. As a result they all show the same time dependence of the generated current for both $x$ and $y$ directions. At the same time, 
for $\mathrm{MoTe_2}$ monolayer, the CB 
population distribution is completely different. While along the direction of the pulse polarization, i.e., $x$ direction, $N_\mathrm{CB}^\mathrm{(res)}\bf(k)$ is concentrated near the $K$ and $K^\prime$ points, in the perpendicular direction, i.e., $y$ direction, it is highly delocalized and there is a large CB population along the lines connecting the $K$ and $K^\prime$ points - see Fig. \ref{Res_CB_TMDCs_Fx_0p25}(c). Thus, along the $x$ direction $N_\mathrm{CB}^\mathrm{(res)}\bf(k)$ behaves similar for $\mathrm{MoTe_2}$ and other TMDC materials and the corresponding current $J_x$ shows similar time dependence for all TMDC monolayers. Along the $y$ direction $N_\mathrm{CB}^\mathrm{(res)}\bf(k)$ of $\mathrm{MoTe_2}$ monolayer is much more extended compared to other TMDC monolayers, as a result the corresponding current, $J_y$,  has completely different time dependence for $\mathrm{MoTe_2}$ monolayer.

\begin{figure}
\begin{center}\includegraphics[width=0.47\textwidth]{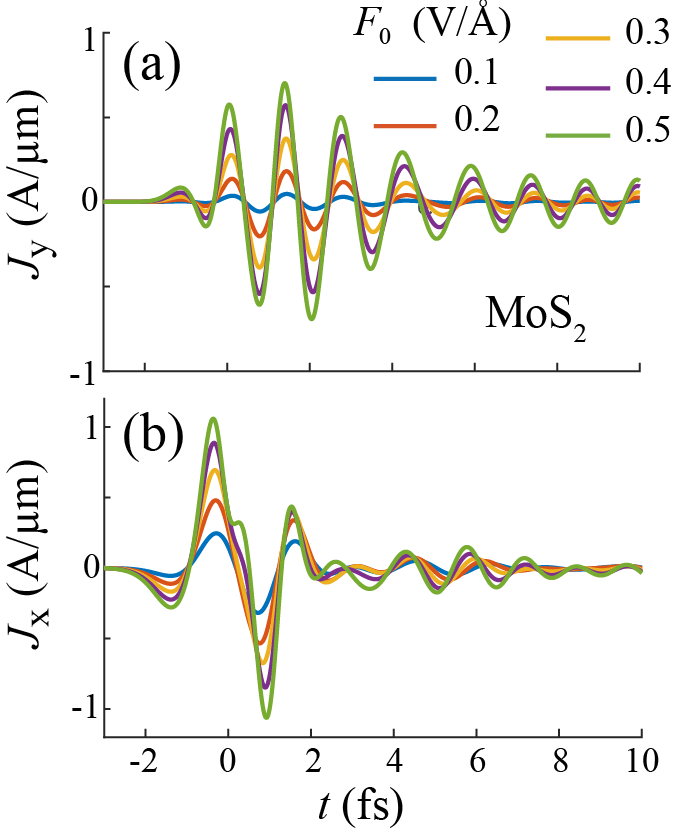}\end{center}
\caption{(Color online) Ultrafast field driven currents in $\mathrm{MoS_2}$ monolayer
as a function of time for different field amplitudes. The $y$ component (a) and the $x$ component (b) of the current are shown. The optical pulse is linearly polarized in the $x$ direction.
} 
\label{J_Fx_F0_MoS2}
\end{figure}

The dependence of the electric current on the field amplitude, $F_0$, is shown in Fig.\ \ref{J_Fx_F0_MoS2} for $\mathrm{MoS_2}$ monolayer. For other TMDC materials the dependence on $F_0$ is similar. As expected, the generated current monotonically increases with $F_0$. In residual current, the frequency of oscillations, which is determined by the bandgap, does not depend on $F_0$.


One of the characteristics of nonlearity of electron response to an ultrashort pulse is a transferred charge through the system during the pulse, which can be also measured experimentally\cite{Yakovlev_2020_Nat_comm_dielectric,
Stockman_et_al_Nat_Phot_2013_CEP_Detector}. The transferred charge is defined by the following expression 
\begin{equation}
\mathbf{Q}=\int_{-\infty}^{\infty } \mathbf{J}(t)\mathrm{d}t^\prime.
\end{equation}
Since the residual current shows an oscillating behavior, to eliminate the dependence on the upper limit in the above integral we introduce a relaxation time of $5$ fs when calculating the transferring charge. The transferred charge is also the residual polarization of the system. 

\begin{figure}
\begin{center}\includegraphics[width=0.48\textwidth]{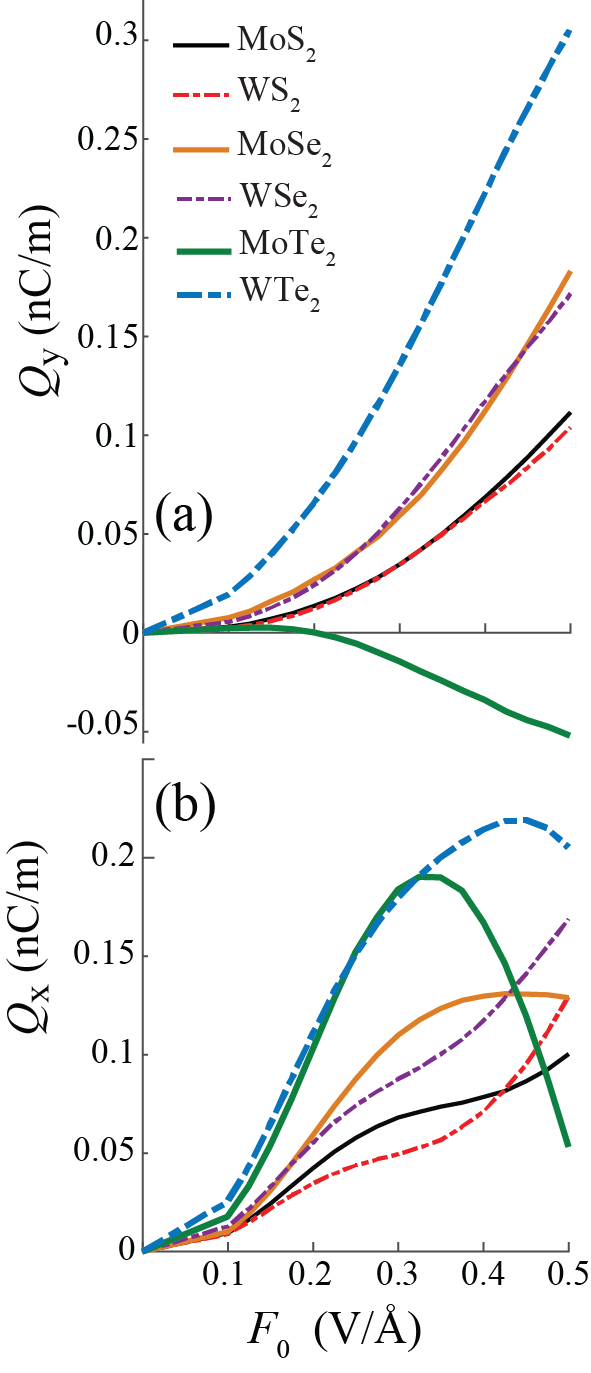}\end{center}
\caption{(Color online) Charge transferred through the system during the pulse as a function of the field amplitude, $F_0$, for different TMDC monolayers. The transferred charge along the $y$ direction (a) and the $x$ direction (b) is shown. The optical pulse is linearly polarized in the $x$ direction.
} 
\label{P_Fx_F0_0p2_TMDC}
\end{figure}

%

The transferred charge as a function of the field amplitude, $F_0$, is shown in Fig. \ref{P_Fx_F0_0p2_TMDC} for different TMDC monolayers. The transferred charge along the $y$ direction, $Q_y$, monotonically increases with $F_0$ - see  Fig. \ref{P_Fx_F0_0p2_TMDC}(a). For all TMDC monolayers, except $\mathrm{MoTe_2}$, the charge is transferred in the positive direction of the $y$ axis, while for $\mathrm{MoTe_2}$ the transfer of the charge occurs in the negative direction of the $y$ axis. Such direction of the transfer corresponds to the condition that the field maximum of the pulse is in the positive direction of the $x$ axis. The magnitude of the transferred charge increases with decreasing the bandgap of TMDC monolayer. The largest charge transfer occurs for $\mathrm{WTe_2}$ monolayer, while the smallest - for $\mathrm{MoTe_2}$ monolayer.


Along the $x$ axis [see Fig.\ \ref{P_Fx_F0_0p2_TMDC}(b)], the charge is transferred in the direction of the field maximum for all TMDC monolayers. The dependence of $Q_x$ on the pulse amplitude, $F_0$, is nonmonotonic. The transferred charge reaches its maximum at some value of $F_0= F_{max}$ and then decreases with $F_0$. The value of $F_{max}$ is partially correlated with the condition that at this field an electron, which is initially at one valley, say valley $K$, reaches the second valley, $K^{\prime }$, during the pulse. For example, for TMDC monolayers with large 
lattice constants, $\mathrm{MoTe_2:~3.557~\AA}$ and  $\mathrm{WTe_2:~3.560~\AA}$\cite{Liu_et_al_PRB_2014_Three_Band_Model}, the maxima occur at lower field amplitudes. Although the lattice constant  is not the only parameter, which determines $Q_x$  dependence on $F_0$, the transferred charge also depends on the bandgap and spin orbit coupling of TMDC monolayer. 
In terms of applications, the data in Fig.\ \ref{P_Fx_F0_0p2_TMDC}(b) illustrate that $\mathrm{MoTe_2}$ monolayer is the most sensitive to the pulse amplitude, i.e., for  $\mathrm{MoTe_2}$ monolayer, the transferred charge, $Q_x$, shows relatively sharp maximum with strong dependence on $F_0$.

\section{Conclusion}

The ultrafast field driven currents in solids are governed by interband and intraband electron dynamics, resulting in two contributions, intraband and interband, to the electric current. In TMDC monolayers, the generated electric current is mainly determined by the interband contribution. As a result, the residual current as a function of time shows oscillations, the frequency of which is determined by the bandgap of the corresponding TMDC monolayer. The TMDC monolayers have broken inversion symmetry, their symmetry group  is $D_{3h}$, and they have only three axises of symmetry, which are along the armchair directions. If the optical pulse is polarized along the direction of symmetry of the monolayer then  the electric current  is generated only along the direction of polarization. But if the polarization of the optical pulse is along a non-symmetric direction, for example, along the zigzag direction, then the electric current 
is generated both along the direction of polarization and in the perpendicular direction. 

For all TMDC monolayers the longitudinal electric current, i.a., the current along the direction of polarization of the pulse, 
show similar behavior as a function of time. Such current transfers an electric charge along the direction of the field maximum. As a function of the field amplitude, the transferred charge has a maximum, the position of which depends on the lattice constant of TMDC monolayer. $\mathrm{MoTe_2}$ monolayer is the most sensitive to the parameters of the optical pulse. 

The transverse current also results in the charge transfer through the system during the pulse. The magnitude of the transferred charge monotonically increases with the field amplitude, while the direction of the transfer depends on the TMDC material. 
Control of an electron transport on a femtosecond time scale pave the way for ultrafast electronic application of TMDCs monolayers.

\appendix
\section{Tight binding Hamiltonian}

The three band  nearest-neighbor (TNN) tight-binding Hamiltonian, $H^\mathrm{({TNN})}$, of TMDC monolayer is constructed from three orbitals (${d_{z^2}}$, ${d_{xy}}$, and ${d_{x^2-y^2}}$) of transition metal atoms \cite{Liu_et_al_PRB_2014_Three_Band_Model}. The Hamiltonian is given by the following expression 
\begin{equation}
H^\mathrm{TNN}(\mathbf k)=\left[ {\begin{array}{ccc}
V_0 &V_1 & V_2\\
V^*_1 &V_{11} & V_{12}\\
V^*_2 &V^*_{12} & V_{22}\
\end{array} } \right]~,
\label{eq:Hamiltonian}
\end{equation}
where 

\begin{eqnarray}
{V_0}&=&{\epsilon_1}+{2t_0 (2 \cos\alpha \cos\beta+\cos2\alpha)}\nonumber\\&+&
{2r_0(2\cos3\alpha \cos\beta+\cos2\beta)}\nonumber\\&+&
{2u_0(2\cos2\alpha \cos2\beta +\cos4\alpha})~,
   \nonumber\\
\mathrm{Re}[V_1]&=&-2 \sqrt{3}t_2\sin\alpha \sin\beta+
   {2(r_1+r_2)\sin3\alpha \sin\beta}\nonumber\\&-&
   {2\sqrt{3}u_2\sin2\alpha \sin2\beta}~,
\nonumber\\
\mathrm{Im}[V_1]&=&2t_1\sin \alpha(2\cos\alpha +\cos \beta)\nonumber\\&+& 2(r_1-r_2)\sin3\alpha \cos\beta\nonumber\\&+& 2u_1\sin2\alpha (2\cos2\alpha+\cos2\beta)~,
 \nonumber\\
\mathrm{Re}[V_2]&=&2t_2(\cos2 \alpha - \cos \alpha \cos \beta)\nonumber\\&-&\frac{2}{\sqrt{3}}(r_1+r_2)(\cos3\alpha \cos\beta - \cos 2\beta)\nonumber\\&+& 2u_2(\cos4\alpha-\cos2\alpha \cos2\beta)~,
 \nonumber\\
\mathrm{Im}[V_2]&=&2\sqrt{3}t_1\cos\alpha \sin\beta\nonumber\\&+&
 \frac{2}{\sqrt{3}}\sin\beta (r_1-r_2)(\cos3\alpha+2\cos\beta)\nonumber\\&+&
2\sqrt{3}u_1\cos2\alpha \sin2\beta~,
 \nonumber\\
V_{11}&=&\epsilon_2+(t_{11}+3t_{22})\cos\alpha \cos\beta\nonumber\\&+&2t_{11}\cos2\alpha+
4r_{11}\cos3\alpha \cos\beta\nonumber\\&+&2(r_{11}+\sqrt{3}r_{12}\cos2\beta)
\nonumber\\
 &+&(u_{11}+3u_{22})\cos2\alpha \cos2\beta +2u_{11}\cos4\alpha~,
 \nonumber\\
\mathrm{Re}[V_{12}]&=&\sqrt{3}(t_{22}-t_{11})\sin\alpha sin\beta \nonumber\\&+&4r_{12}\sin3\alpha \sin \beta \nonumber\\&+&  \sqrt{3}(u_{22}-u_{11}\sin2\alpha \sin2\beta)~,
\nonumber\\
\mathrm{Im}[V_{12}]&=&4t_{12}\sin\alpha(\cos\alpha-cos\beta)\nonumber\\&+&4u_{12}\sin2\alpha(\cos2\alpha-\cos2\beta)~,
\nonumber\\
V_{22}&=&\epsilon_2+(3t_{11}+t_{22})\cos\alpha \cos \beta +2t_{22}\cos2\alpha  \nonumber\\&+&2r_{11}(2\cos3\alpha \cos \beta +\cos 2\beta)
\nonumber\\
&+&\frac{2}{\sqrt{3}}r_{12}(4\cos3\alpha \cos \beta - \cos 2\beta)\nonumber\\&+&
(3u_{11}+u_{22})\cos2\alpha \cos2\beta +2u_{22}\cos 4\alpha~,
\nonumber\\
\end{eqnarray}
and
\begin{equation}
{(\alpha,\beta)=\left(\frac{1}{2}k_xa,\frac{\sqrt{3}}{2}k_ya\right)}~.
\end{equation}

The values of the parameters in the above Hamiltonian 
for different TMDC materials are given in table \ref{T1} Ref [\onlinecite{Liu_et_al_PRB_2014_Three_Band_Model}].

\begin{table}
\begin{center}
\begin{tabular}{|c|c|c|c|c|c|c| }
\hline
\multirow{1}{2.5em}{}&$\mathrm{MoS_2}$ & $\mathrm{WS_2}$ & $\mathrm{MoSe_2}$ & $\mathrm{WSe_2}$ &$\mathrm{MoTe_2}$ & $\mathrm{WTe_2}$\\
\hline
\multirow{1}{2.5em}{a} &3.19&3.191&3.326&3.325&3.557&3.560 \\
\hline
\multirow{1}{2.5em}{$\epsilon_1$} &0.683&0.717&0.684&0.728&0.588&0.697\\
 \hline
\multirow{1}{2.5em}{$\epsilon_2$} &1.707&1.916&1.546&1.655&1.303&1.380\\
 \hline
\multirow{1}{2.5em}{$t_0$} &-0.146&-0.152&-0.146&-0.146&-0.226&-0.109\\
 \hline
\multirow{1}{2.5em}{$t_1$} &-0.114&-0.097&-0.130&-0.124&-0.234&-0.164\\
 \hline
\multirow{1}{2.5em}{$t_2$} &0.506&0.590&0.432&0.507&0.036&0.368\\
 \hline
\multirow{1}{2.5em}{$t_{11}$} &0.085&0.047&0.144&0.117&0.400&0.204\\
 \hline
\multirow{1}{2.5em}{$t_{12}$} &0.162&0.178&0.117&0.127& 0.098& 0.093\\
 \hline
\multirow{1}{2.5em}{$t_{22}$} &0.073&0.016&0.075&0.015&0.017&0.038\\
 \hline
\multirow{1}{2.5em}{$r_0$} &0.06& 0.069&0.039&0.036& 0.003&-0.015\\
 \hline
 \multirow{1}{2.5em}{${r_1}$} &-0.236&-0.261&-0.209&-0.234&-0.025&-0.209\\
 \hline
\multirow{1}{2.5em}{$r_{11}$} &0.016&-0.003&0.052& 0.044& 0.082&0.115\\
 \hline
\multirow{1}{2.5em}{$r_{12}$} &0.087&0.109&0.060&0.075&0.051&0.009\\
\hline
\multirow{1}{2.5em}{$r_2$} &0.067&0.107&0.069&0.107&-0.169&0.107\\
 \hline
\multirow{1}{2.5em}{$u_{0}$} &-0.038&-0.054&-0.042&-0.061&0.057&-0.066\\
 \hline
\multirow{1}{2.5em}{$u_{1}$} &0.046&0.045&0.036&0.032&0.103&0.011\\
 \hline
\multirow{1}{2.5em}{$u_{2}$} &0.001&0.002&0.008&0.007&0.187&-0.013\\
 \hline
\multirow{1}{2.5em}{$u_{11}$} &0.266&0.325&0.272&0.329&-0.045&0.312\\
 \hline
\multirow{1}{2.5em}{$u_{12}$} &-0.176&-0.206&-0.172&-0.202&-0.141&-0.177\\
 \hline
\multirow{1}{2.5em}{$u_{22}$} &-0.15&-0.163&-0.150&-0.164&0.087&-0.132\\
 \hline
\multirow{1}{2.5em}{$\lambda$} &0.073&0.211& 0.091& 0.228&0.107& 0.237\\
 \hline
\end{tabular}

\end{center}
\caption{Parameters of three band tight-binding Hamiltonian. Here the lattice constant, $a$, is in units of $\mathrm{\AA}$, while all other parameters are in units of eV\cite{Liu_et_al_PRB_2014_Three_Band_Model}.}
\label{T1}
\end{table}

---------------------

\begin{acknowledgments}
Major funding was provided by Grant No. DE-FG02-01ER15213
from the Chemical Sciences, Biosciences and Geosciences
Division, Office of Basic Energy Sciences, Office of Science,
US Department of Energy.
Numerical simulations were performed using support by Grant No. DE-SC0007043
from the Materials Sciences and Engineering Division of
the Office of the Basic Energy Sciences, Office of Science,
US Department of Energy. 
\end{acknowledgments}

%

\end{document}